\documentclass[12pt]{article}

\usepackage{amsmath}
\usepackage{amsfonts}

\newtheorem{theorem}{Theorem}

\newtheorem{remark}[theorem]{Remark}
\newtheorem{proposition}[theorem]{Proposition}

\begin{document}

\title{The Arrhenius formula in kinetic theory \\ and Witten's spectral asymptotics}

\author{S.V.Kozyrev\footnote{Steklov Mathematical Institute, Gubkina str. 8, 119991, Moscow, Russia}, I.V.Volovich\footnote{Steklov Mathematical Institute, Gubkina str. 8, 119991, Moscow, Russia}}

\maketitle

\begin{abstract}
A new approach to the proof of the Arrhenius formula of kinetic theory is proposed. We prove this formula starting from the equation of diffusion in a potential. We put this diffusion equation in the form of evolutionary equation generated by some Schr\"odinger operator.
We show that the Arrhenius formula for the rate of over the barrier transitions follows from the formula for the rate of quantum tunnel transitions for the considered Schr\"odinger operator.

Relation of the proposed approach and the Witten method of the proof of the Morse inequalities is discussed. In our approach the Witten spectral asymptotics takes the form of the low temperature limit and the Arrhenius formula is a correction to the Witten asymptotics.
\end{abstract}

\section{Introduction}

The Arrhenius formula \cite{Berlin}  for the reaction rate in chemical kinetics describes chemical reactions which are related to the over barrier transitions  between the potential wells. With this formula the reaction rate is proportional to
$$
e^{-\beta \Delta E}
$$
where $\Delta E$ is the difference of the energies of the transition state (the activation barrier) and of the initial state.

The Eyring formula for the reaction rate (the modification of the Arrhenius formula)
\begin{equation}
\label{Arrhenius} e^{-\beta \Delta F}
\end{equation}
replaces the difference of energies by the difference of the free energies $\Delta F$.

Let us remind that the free energy
$F(G)$ of the set of states $G$ is defined as:
$$
e^{-\beta F(G)}=\int_G e^{-\beta U(x)}dx.
$$
Here $U(x)$ is the energy of the state $x$.

In the present paper we discuss the proof of the Arrhenius and the Eyring formulas starting from the equation of diffusion in the potential
\begin{equation}\label{diff} {\partial f\over \partial t}=\Delta
f+\beta \nabla f \cdot \nabla U+\beta f \Delta U,
\end{equation}
where $x\in\mathbb{R}^d$, $f=f(x,t)$
is the distribution function, $U=U(x)$ is the potential (a real valued function), $U\in C^2(\mathbb{R}^d)$,
$\beta=1/kT>0$ is the inverse temperature.

We show that the above equation is equivalent to the evolution equation generated by some Schr\"odinger operator. The ground state for this operator corresponds to the Gibbs state for the diffusion in the potential (\ref{diff}). The energy of the first excited state describes the velocity of relaxation to the Gibbs state.

For the case of the double well potential the energy of the first excited state can be estimated quasiclassically. In the quasiclassical approach the splitting of the ground state for the Schr\"odinger operator is generated by the tunnel transitions between the potential wells.
We show that this estimate for the energy of the first excited state gives the Arrhenius formula.

Therefore we prove the Arrhenius formula which describes the over barrier transitions in the classical kinetic theory using the spectral properties of some Schr\"odinger operator related to quantum tunneling under barrier transitions.

The relation of the Arrhenius formula and non--Arrhenius relaxation on complex landscapes of energy was discussed in many works, in particular, see \cite{Berlin}. In \cite{ABK} the relation of dynamics on complex landscapes and $p$-adic diffusion was considered. In \cite{Goldansky}, \cite{Goldansky1} the quantum low temperature limit for the reaction rates which is related to tunneling transitions was discussed, see also \cite{QuantumRate}  for the discussion of the quantum corrections to the Arrhenius formula.  In the present paper we use quantum tunneling for the corresponding Schr\"odinger operator to prove the purely classical Arrhenius formula.

The structure of the present paper is as follows.

In section 2 we discuss the diffusion equation in the potential and construct the corresponding Schr\"odinger operator. We also prove the stabilization property for the evolution described by the considered diffusion equation.

In section 3 the quasiclassical approach to the introduced in section 2 Schr\"odinger operator is discussed. We show that in the regime of low temperatures the bound states correspond to potential wells related to local minima of the potential $U$ (i.e. the potential in the diffusion equation (\ref{diff})). We prove that the quasiclassical formula for the splitting of the ground state for the double well potential in the low temperature limit implies the Arrhenius and Eyring formulas.

In section 4 we prove the Eyring formula beyond the frameworks of the quasiclassical approximation.

In section 5 we discuss the relation of the approach of the present paper and the Witten approach in the Morse theory.

\section{The diffusion equation and the Schr\"odinger operator}

Equation (\ref{diff}) can be put in the form
$$
{\partial f\over \partial t}={\rm\bf div}\,\left[e^{-\beta U}\,{\rm\bf grad}\,
\left[f e^{\beta U}\right]\right].
$$
The equilibrium state $f(x)={\rm const}\,e^{-\beta U(x)}$
(the Gibbs distribution) is the stationary solution of the above equation.

With the substitution $\psi=e^{\beta U/2}f$ this equation is put in the form
\begin{equation}\label{WickSchro} {\partial\over \partial t}\psi=-H\psi,\qquad H= -\left[e^{\beta
U/2}\nabla e^{-\beta U/2}\right]\cdot \left[e^{-\beta U/2}\nabla
e^{\beta U/2}\right].
\end{equation}

The operator $H$ is self--adjoint and positive, since
$H$ is a finite sum of positive operators
\begin{equation}\label{AstarA}
H=A^*\cdot A=\sum_{i=1}^d A_i^*A_i,\qquad A=e^{-\beta
U/2}\nabla e^{\beta U/2},\quad A_i=e^{-\beta
U/2}{\partial\over\partial x_i} e^{\beta U/2}.
\end{equation}

The gradient part of $H$ cancels and we get the Schr\"odinger operator
\begin{equation}\label{H}
H=-\Delta+V,\qquad V=-{\beta\over 2}\Delta
U+{\beta^2\over 4}(\nabla U)^2 .
\end{equation}
Here $V$ is the effective potential. The spectrum of the Schr\"odinger operator $H$ is bounded from below by zero. Since the function $e^{-\beta U/2}$ satisfies
$$
He^{-\beta U/2}=0,
$$
then, if $e^{-\beta U/2}\in L^2(\mathbb{R}^d)$, i.e. the Gibbs state is integrable
$e^{-\beta U}\in L^1(\mathbb{R}^d)$, zero will be the minimal point of the discrete spectrum of $H$ and $e^{-\beta U/2}$ will be
the ground state.

Let also the operator $H$ satisfies the condition of the existence of the energy gap: the spectrum of $H$ is a subset of $\{0\}\bigcup [E_1,\infty)$, $E_1>0$ (we assume that $E_1$ belongs to the spectrum). The value $E_1$ we call the energy gap for the operator $H$.

The positivity of $H$ and the existence of the energy gap implies the convergence of the solution of equation (\ref{diff}) to the Gibbs state with $t\to\infty$. We formulate the following theorem  (a variant of a theorem about stabilization of solutions of parabolic equations).

\begin{theorem} {\sl Let for equation (\ref{diff}) the potential satisfies  $U\in C^2(\mathbb{R}^d)$ and the Gibbs state is integrable
$$
\int e^{-\beta U(x)}dx<\infty.
$$

Let also the potential $U$ satisfies the condition of the existence of energy gap.

Consider for equation (\ref{diff})
a Cauchy problem in the space $L^2(\mathbb{R}^d,e^{\beta U(x)}dx)$ of quadratically integrable with the weight $e^{\beta U}$ functions. The initial condition $f_0$ for a Cauchy problem belongs to this space, i.e.  $f_0$ satisfies
$$
\int |f_0(x)|^2e^{\beta
U(x)}dx<\infty.
$$

Then the solution $f(t,x)$ of the described Cauchy problem for equation (\ref{diff}) with the initial condition $f_0(x)$ exists, is unique and with $t\to\infty$ tends to the function proportional to the Gibbs state
$$
\lim_{t\to\infty}f(x,t)=e^{-\beta U(x)}{\int f_0(x)dx\over \int
e^{-\beta U(x)}dx}.
$$

}

\end{theorem}

\noindent{\it Proof}\quad The scalar product of the functions $\psi(x,t)=f(x,t)e^{\beta U(x)/2}$ and $\psi_0(x)=e^{-\beta U(x)/2}$ in $L^2(\mathbb{R}^d)$ is
$$
\langle \psi,\psi_0\rangle_{L^2(\mathbb{R}^d)}=\int f(x,t)dx.
$$
This implies that the solution $f$ of (\ref{diff}) satisfies the condition of conservation of the number of particles
\begin{equation}\label{number}
{d\over dt}\int f(x,t)dx=0
\end{equation}
because the energy of the ground state $\psi_0$ is equal to zero.

Then since the energy of the ground state is zero and there exists the energy gap
$$
\lim_{t\to\infty}f(x,t)=\lim_{t\to\infty}\psi(x,t)e^{-\beta U(x)/2}=\lim_{t\to\infty} \langle \psi,\psi_0\rangle_{L^2(\mathbb{R}^d)}\psi_0(x)e^{-\beta U(x)/2}/\|\psi_0\|^2.
$$
By (\ref{number}) this implies
$$
\lim_{t\to\infty}f(x,t)=e^{-\beta U(x)}{\int f_0(x)dx\over \int
e^{-\beta U(x)}dx}.
$$
which finishes the proof of the theorem. $\square$

\bigskip

Let us discuss the behavior of the solution for the case when the operator $H$ has the discrete spectrum and the initial condition $\psi(0)$ for (\ref{WickSchro}) belongs to $L^2(\mathbb{R}^d)$. Let the initial condition possesses the expansion $\psi(0)=\sum_{i=0}^{\infty}\psi_i$ over the eigenvectors of $H$, which correspond to the eigenvalues $E_i$, $E_i<E_{i+1}$, $\psi_0$ is proportional to $e^{-\beta U/2}$, $E_0=0$.
Then
$$
\psi(t)=\psi_0+\psi_1 e^{-E_1t}+\sum_{i=2}^{\infty}\psi_ie^{-E_it}.
$$
Therefore the evolution $\psi(t)$ tends to the ground state $\psi_0$ proportional to $e^{-\beta U/2}$, and the rate of the convergence for large $t$ is described by the term $\psi_1 e^{-E_1t}$ in the above expansion corresponding to the first excited level of the Hamiltonian $H$.

\bigskip

\noindent{\bf Example}\quad
Let us discuss the case of the quadratic potential $U=\alpha |x|^2$. Then the effective Schr\"odinger potential $V$ is also quadratic:
$$
V=-{\beta\over 2}\Delta U+{\beta^2\over 4}(\nabla U)^2=-\beta\alpha
d+\beta^2\alpha^2|x|^2.
$$
Here $d$ is the dimension of the space.

$$
H=-\Delta +V=\sum_{i=1}^d 2\beta\alpha{1\over 2}\left[-{1\over
\beta\alpha}{\partial^2\over\partial x_i^2}+\beta\alpha x_i^2-1\right].
$$

We have $d$ harmonic oscillators with the eigenvalues
$E_i=2\beta\alpha n_i$, $n_i=0$, 1, 2, \dots. In particular the term
$-{\beta\over 2}\Delta U$ cancels the energy of the vacuum and makes the energy of the ground state for $H$ equal to zero.

The energy gap (between the ground and the first excited states) in this model is equal to $2\beta\alpha$. In the considered single well case the Arrhenius formula is not applicable.

\begin{remark}\label{minima}\quad{\rm
We consider the effective potential $V=-{\beta\over 2}\Delta U+{\beta^2\over 4}(\nabla U)^2$ in the regime of low temperatures (large $\beta$) and a generic potential $U$. Generic here means that $U$ is a Morse function i.e. it possesses a finite number of critical points (the points where $\nabla U=0$) and the Hessian (the matrix of second derivatives of $U$) in the critical points is non--degenerate.

Since in the low temperature regime the leading contribution to the effective potential comes from the term containing the square of the gradient of $U$, the minima of the effective potential $V$ will correspond to the critical points of the potential $U$. In this case the values of the effective potential $V$ in the minima corresponding to the minima of $U$ will be negative (by the positivity of the second derivatives in the minima of $U$). In the minima of $V$ corresponding to the maxima of $U$ the values of $V$ will be positive. For saddle points of $U$ the values of $V$ in the corresponding minima will depend on the sign of the Laplacian of $U$.

In the next section we show that the minima of the potential $U$ satisfy the Bohr--Sommerfeld condition of existence of bound states for $V$.
}
\end{remark}

\section{Quasiclassical approach}

In the present section we consider the quasiclassical approximation for the Schr\"odinger operator under consideration (in this section we restrict ourselves to the one dimensional case and the double well potential). We show that in the low temperature regime in this approximation there exist states of the discrete spectrum for the wells of the effective potential $V$ which correspond to the local minima of $U$ and the splitting of  energy for the ground state related to transitions between these potential wells is described by the Arrhenius formula.

\begin{remark}{\rm
The quasiclassical Bohr--Sommerfeld quantization condition has the form \cite{Landau}
$$
{1\over 2\pi\hbar}\int p(x)dx=n+{1\over 2},
$$
where the integration runs over the periodic orbit of the quasiclassical particle, $n=0,1,2,\dots$ is the number of the quantum state.

Let us investigate, using this condition of possibility of existence of a bound quantum state in the potential well for operator (\ref{H}).
For the ground state we have the energy $E=0$, the momentum of the quantum particle $|p|=\sqrt{2m(E-V)}=\sqrt{-2mV}$ with the normalization $\hbar=2m=1$.

We assume that the bound state in the potential well exists if the Bohr--Sommerfeld condition can be satisfied for $n=0$, i.e.
\begin{equation}\label{bohr_sommerfeld}
{1\over \pi}\int_{-V\ge 0} \sqrt{-V(x)}dx\ge {1\over 2},
\end{equation}
where the integration runs over the domain where the potential $V=-{\beta\over 2}\Delta
U+{\beta^2\over 4}(\nabla U)^2$ is non--positive.

Minima of the potential $V$ are described in remark \ref{minima}. For large $\beta$ (low temperatures) minima of $V$ correspond to critical points of $U$. Minima of $V$ related to maxima of $U$ do not satisfy the condition (\ref{bohr_sommerfeld}) since for these minima the potential $V$ is positive.

For minima of $V$ corresponding to the minima of $U$ the potential $V$ is negative. We get in the one--dimensional case, expanding the potential $U$ in the vicinity of the minimum $x_0$
\begin{equation}\label{estimate}
{1\over \pi}\int_{-V\ge 0} \sqrt{{\beta\over 2}U''(x_0)-{\beta^2\over 4}(U''(x_0)x)^2}dx
= {1\over \pi}\int_{-1}^{1} \sqrt{1-x^2}dx={1\over 2}.
\end{equation}
Therefore for these minima of $V$ the condition (\ref{bohr_sommerfeld}) will be satisfied.

In the following we will ignore the potential wells of $V$ corresponding to the critical points of $U$ which are not local minima.

}
\end{remark}

Let us consider the one dimensional double well potential which contains two symmetric potential wells separated by the potential barrier. Tunnel transitions between the potential wells imply the splitting of the energy of the ground state i.e. the ground state becomes the pair of close energy levels and the difference of the corresponding energies is related to the rate of tunnel transitions between the potential wells.

For the dynamics generated by the Schr\"odinger operator with this potential the rate of relaxation to the ground state for large times will be described by the energy $E_1$ of the first excited state i.e. by the splitting of energy of the ground state due to the tunnel transitions. Let us show that for the Schr\"odinger operator under consideration this energy splitting will be described by the Arrhenius formula.

\medskip

The formula for the splitting of the energy of the ground state in the symmetric one dimensional double well potential in the quasiclassical approximation \cite{Landau}, Chapter 7, section 50, problem 3, has the form
$$
E_1-E_0={|p(0)|\over {\rm Vol}}\exp\left(-{1\over\hbar}\int_{-a}^{a}|p(x)|dx\right).
$$
Here $p(0)$ is the momentum of the particle at the barrier of energy and ${\rm Vol}$ is the volume of one of the potential wells.
The integration runs over the interval of the tunnel transition where the potential $V$ is positive i.e.
${\beta\over 2}{U'}^2> U''$. The momentum of the quantum particle $|p|=\sqrt{2m(V-E_0)}=\sqrt{2mV}$ with the normalization used $\hbar=2m=1$. Since $E_0=0$ we have $E_1-E_0=E_1$.

We get
$$
E_1={|p(0)|\over {\rm Vol}}\exp\left(-\int_{-a}^{a}\sqrt{V(x)}dx\right) ={|p(0)|\over {\rm Vol}}\exp\left(-\int_{{\beta\over 2}{U'}^2>
U''}\sqrt{-{\beta\over 2} U''(x)+{\beta^2\over 4}{U'}^2(x)}dx\right).
$$

We are interested in the low temperature limit when the transitions between the potential wells are slow. In this case the inverse temperature  $\beta$ is large and we can omit the first term under the square root in  the expression for $E_1$ and obtain the expression
$$
{|p(0)|\over {\rm Vol}}\exp\left(-{\beta\over 2}\int_{{\beta\over 2}{U'}^2>
U''}|U'(x)|dx\right).
$$

Integration runs over the symmetric domain of the potential barrier, therefore
$$
\int_{{\beta\over 2}{U'}^2> U''}|U'(x)|dx=2\int_{{\beta\over 2}{U'}^2> U'';x\le 0}U'(x)dx=2\Delta U,
$$
where the energy barrier $\Delta U$ is equal to the difference of values of $U$ at the potential barrier (for $x=0$) and in the well (at the point where ${\beta\over 2}{U'}^2= U''$). Since the inverse temperature $\beta$ is large the condition ${\beta\over 2}{U'}^2= U''$ is approximately satisfied at the critical point $U'=0$.

This implies that
$$
E_1={|p(0)|\over {\rm Vol}}\exp\left(-\beta  \Delta U \right),
$$
i.e. the energy splitting of the ground state for the operator (\ref{H}) and correspondingly the relaxation rate for equation (\ref{diff}) possess the Arrhenius dependence.

Taking into account the multiplier ${|p(0)|/ {\rm Vol}}$ we can get the Eyring formula. We have
$$
|p(0)|=\sqrt{-{\beta\over 2}U''(x_0)}
$$
where $x_0$ is at the energy barrier. Modeling the potential in the vicinity of the energy barrier by the Gaussian expression $\Delta U\exp(-a^2(x-x_0)^2)$ we get
$|p|=a\sqrt{\beta \Delta U}$ and
$$
E_1=\sqrt{\beta \Delta U}{a\over {\rm Vol}}\exp\left(-\beta  \Delta U \right).
$$
Since $\ln a$ is the entropy of the transition state and $\ln\,{\rm Vol}$ is the entropy of the initial state (the potential well), the above formula for the energy of the first excited state for the double well potential coincides with the Eyring formula (one can ignore the slow dependence on $\sqrt{\beta \Delta U}$).

\section{The Eyring formula}

In the present section we study the splitting of the energy levels for the Schr\"odinger operator (\ref{H}) in the case of the double well potential $U$ in the multidimensional case. Since we know the exact expression for the ground state we can perform the computations analogous to the computations made in the previous section beyond the frameworks of the quasiclassical approximation.

Let us consider the normalized ground state (with the norm one)
$$
\psi_0(x)={e^{-{\beta\over 2}U(x)}\over \left[\int e^{-\beta
U(x)}dx\right]^{1\over 2}}
$$
and the normalized first excited state $\psi_1$ for the Schr\"odinger operator (\ref{H}). These states satisfy the equations
\begin{equation}\label{psi0}
H\psi_0=\left(-\Delta-{\beta\over 2}\Delta U+{\beta^2\over 4}(\nabla
U)^2\right)\psi_0=0,
\end{equation}
\begin{equation}\label{psi1}
H\psi_1=\left(-\Delta-{\beta\over 2}\Delta U+{\beta^2\over 4}(\nabla
U)^2\right)\psi_1=E_1 \psi_1.
\end{equation}

Let us multiply the first equation by $\psi_1$ and the second equation by $\psi_0$ and subtract the second equation from the first one:
$$
\psi_1\Delta\psi_0-\psi_0\Delta\psi_1=E_1 \psi_1\psi_0.
$$

Let us construct a surface $S$ which goes through the saddle point (the transition state between the two potential wells) and separates all the space in the two domains which contain the corresponding potential wells of $U$. Namely the surface $S$ goes through the saddle point and the gradient $\nabla U$ is tangent to $S$ (equivalently is orthogonal to the normal to $S$)\footnote{For the dimension of the space larger than two such a surface is not unique. We choose some surface of this kind.}.

We integrate over one of the domains $G$ separated by $S$ (say over the domain which contains the first potential well). Applying the Green formula we get
$$
\int_{G}(\psi_1\Delta \psi_0-\psi_0\Delta \psi_1)dx=\int_{\partial
G}\left(\psi_1{\partial \psi_0\over \partial n}-\psi_0{\partial
\psi_1\over
\partial n}\right)dS=E_1\int_{G} \psi_1\psi_0dx.
$$
Since $\psi_0$ is proportional to $e^{-{\beta\over 2}U}$ the derivative of $\psi_0$ with respect to the normal to $S$ is equal to zero: ${\partial
\psi_0\over \partial n}=0$.

We get the following (exact) expression of the energy of the first excited state using the wave functions of the ground and the first excited states
\begin{equation}\label{E}
E_1=-{\int_{\partial G}\psi_0{\partial \psi_1\over
\partial n}dS\over\int_{G} \psi_1\psi_0dx}.
\end{equation}

Expression (\ref{E}) for the energy of the first excited state contains the integral in the denominator which can be estimated as
$$
\int_{G} \psi_1\psi_0dx=\int_{G} \psi^2_0dx=\int_{G} e^{-\beta U(x)}dx=e^{-\beta F_0},
$$
since the ground and the first excited states in one of the potential wells can be considered approximately equal. Here $F_0$ is the free energy of the domain $G$ (the main contribution to the free energy for large $\beta$ comes from the potential well).

Let us discuss the form of the state $\psi_1$. We make in (\ref{psi1}) the substitution $\psi_1(x)=\theta(x)\psi_0(x)$. The function $\theta$ should be close to constants in the potential wells and should have a large gradient in the transition states --- barriers between the potential wells. We get (taking into account that $\psi_0$ is proportional to $e^{-\beta U/2}$) the equation
$$
-\Delta \theta+\beta\nabla \theta\cdot\nabla U =E_1 \theta.
$$

In the one dimensional case this equation reduces to
$$
-\theta''+\beta \theta' U' =E_1 \theta,
$$
where $E_1$ is a small parameter. Omitting the small right hand side of the above expression we get the approximate solution
$$
\theta'(x)=C e^{\beta U(x)}
$$

We choose the normalization constant $C$ using the following statement: the function $\theta$ changes from one to minus one when moving from one potential well to the other. This implies
$$
\theta'(x)={ 2e^{\beta U(x)} \over \int e^{\beta U(x)} dx},
$$
where the integration runs from one potential well to the other. In the following we will use the above approximation for the multidimensional case.

The integral over the surface $S=\partial G$ in the numerator can be estimated as follows. The expression under the integral is approximated by
$$
\psi_0(x){\partial \psi_1(x)\over
\partial n}=-\theta'(x)\psi_0^2(x)=-{ 2e^{\beta U(x)} \over \int e^{\beta U(x)} dx}\psi^2_0=-{ 2 \over \int e^{\beta U(x)} dx}.
$$
The sign changes since the normal to $S$ is directed to the opposite direction. In the following we ignore the coefficient two in the above expression (since this coefficient is beyond our approximation).

For the transition state the integral in the numerator of (\ref{E}) is supported in the vicinity of the energy barrier (where the function $\theta'$ is supported). We get the approximation
$$
\int_{\partial G} \left[\int e^{\beta U(x)} dx\right]^{-1}dS \sim \int_{\partial G} \int e^{-\beta U(x)} dxdS=e^{-\beta F_1},
$$
where $F_1$ is the free energy of the transition state.

The above discussion implies the following:

\begin{proposition}
{\sl The splitting $E_1$ of the ground state for the Schr\"odinger operator (\ref{H}) for the double well potential is approximated by the Eyring formula
$$
E_1=\exp(-\beta(F_1-F_0)).
$$
Here $F_0$ is the free energy of the potential well (the initial state of the corresponding kinetics) and $F_1$ is the free energy of the transition state.
}
\end{proposition}

We proved (at the physical level of rigour) that the transition rate for the kinetics related to the diffusion equation (\ref{diff}) is described by the Eyring formula. In our approach the rate of over barrier transitions is described by the spectral theory of the corresponding Schr\"odinger operator which itself is related to the rate of quantum tunneling transitions between the potential wells.

\section{The Witten Laplacian and the Morse theory}

The Witten approach to the proof of the Morse inequalities Морса \cite{Witten0}, see also \cite{Bismut}, \cite{Witten}, Chapter 11, is related to computation of the spectral asymptotics of the deformed Witten Laplacian  $L_t$,
$$
L_t=d^*_td_t+d_td^*_t,\qquad d_t=e^{-tf}de^{tf},\qquad d^*_t=e^{tf}d^*e^{-tf}.
$$
Here $t\in\mathbb{R}$, $f$ is some Morse function, i.e. a smooth function with the finite number of critical points (points where the gradient is equal to zero) where the Hessian (the form of second derivatives) is non--degenerate, $d$ is the exterior derivative (acting on differential forms), $d^*$ is the Hodge conjugate to $d$, $t\in\mathbb{R}$.

The asymptotics of the spectrum of the deformed Laplacian is computed quasiclassically in the limit $t\to\infty$. The corresponding estimates for the Betti numbers coincide with the Morse inequalities.

The Witten approach is connected to the discussed in the present paper spectral properties of the Schr\"odinger operator in the following way. The Witten asymptotics is the low temperature asymptotics and the Arrhenius formula is a correction to the Witten spectral asymptotics.

The operator $A$ of the form (\ref{AstarA}) is a restriction of the deformed exterior derivative $d_t$ to functions  (i.e. zero--forms). We have $f=U$, $t=\beta/2$. The operator (\ref{H}) takes the form $H=d^*_td_t$ of one of the contributions to the deformed Laplacian, the asymptotics $t\to \infty$ is the low temperature limit.

The Witten approach discusses the spectra of the oscillators related to all critical point  of $f$. In the present paper we consider only the minima of $U$ (not all critical points). The discussed here splitting of the ground state due to tunnel transitions between the potential wells (described by the Arrhenius formula) is a correction to the Witten spectral asymptotics (and vanishes in the proof of the Morse inequalities).

\bigskip

\noindent{\bf Acknowledgments}\qquad The authors would
like to thank A.I.Mikhailov for fruitful discussions. The authors gratefully
acknowledge being partially supported by the grant DFG Project 436 RUS 113/951, by the grants of
the Russian Foundation for Basic Research
RFFI 08-01-00727-a and RFFI 09-01-12161-ofi-m, by the grant of the President of Russian
Federation for the support of scientific schools NSh-7675.2010.1 and
by the Program of the Department of Mathematics of the Russian
Academy of Science ''Modern problems of theoretical mathematics'',
and by the program of Ministry of Education and Science of Russia
''Development of the scientific potential of High School, years of
2009--2010'', project 3341. One of the authors (S.K.) was also partially supported by the grant
DFG Project 436 RUS 113/809/0-1.

\end{document}